\documentclass[aps,prb,twocolumn,groupedaddress,showpacs,showkeys]{revtex4}

\newcommand {\mofe} {\{$\textrm{Mo}_{72}\textrm{Fe}_{30}$\}}
\newcommand {\subhm} {$\mathcal{H}\left(M\right)$\ }
\newcommand {\subhs} {$\mathcal{H}\left(S\right)$\ }

\newcommand{\eqcomma}{\;\;,}

\newcommand{\op}[1]{%
    \fontdimen12\textfont3=2pt\fontdimen12\scriptfont3=1.4pt%
    \!\null\mathop{\vphantom{#1}\smash{#1}}\limits_{\sim}\null\!}
\newcommand{\fmref}[1]{(\protect\ref{#1})}

\usepackage{graphicx}

\begin{document}

\title{Evaluation of the low-lying energy spectrum of magnetic
  Keplerate molecules with DMRG}

\author{Matthias Exler}
\email{matexler@uos.de}
\affiliation{Universit\"at Osnabr\"uck, Fachbereich Physik,
D-49069 Osnabr\"uck}

\author{J\"urgen Schnack}
\email{jschnack@uos.de}
\affiliation{Universit\"at Osnabr\"uck, Fachbereich Physik,
D-49069 Osnabr\"uck}

\date{\today}

\begin{abstract}
  
We apply the density-matrix renormalization group technique to
magnetic molecules in order to evaluate the low-lying energy
spectrum. In particular, we investigate the giant Keplerate molecule
\mofe,\cite{MSS:ACIE99} where 30 $\textrm{Fe}^{3+}$ ions (spins 5/2)
occupy the sites of an icosidodecahedron and interact via
nearest-neighbor antiferromagnetic Heisenberg exchange.

The aim of our investigation is to verify the applicability and
feasibility of DMRG calculations for complex magnetic
molecules. To this end we first use a fictitious 
molecule with the same structure as \mofe\ but with spins 1/2 as a test
system. Here we investigate the accuracy of our DMRG implementation in
comparison to numerically exact results.\cite{SSR:EPJB01} Then we apply
the algorithm to \mofe\ and calculate an approximation of the lowest
energy levels in the subspaces of total magnetic quantum
number. The results prove the existence of 
a lowest rotational band, which was predicted in
Ref.~\onlinecite{LSM:EPL01}.
\end{abstract}

\maketitle

\section{Introduction}

\begin{figure}[ht!]
  \begin{center}
    \scalebox{0.4}{
      \includegraphics[clip]{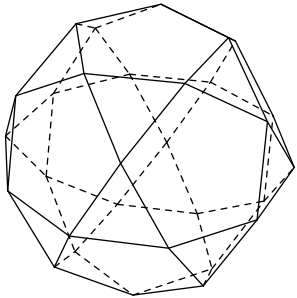}
      }
    \caption{\label{icosidodecahedron-3d.eps}Three-dimensional model
of \mofe: the vertices mark the sites of the $\textrm{Fe}^{3+}$
ions (spin $5/2$), and the lines denote nearest-neighbor interactions.}
  \end{center}
\end{figure}

The rapid progress in polyoxometalate chemistry generates larger
and larger magnetic molecules. The most prominent example of
recent times is the molecular magnet
\mofe, where 30 Fe$^{3+}$ paramagnetic ions
(spin $5/2$)  
occupy the sites of an icosidodecahedron, see
Fig.~\ref{icosidodecahedron-3d.eps}, and interact via 
isotropic nearest-neighbor anti-ferromagnetic Heisenberg
exchange.\cite{MSS:ACIE99,MLS:CPC} Whereas the statistical and
dynamical properties of smaller molecules can be evaluated by
numerically exact diagonalization, the huge dimension of the
Hilbert space of magnetic macromolecules prohibits such
attempts. In the case of \mofe\
this dimension amounts to $6^{30}$, which is beyond the
power of any computer.

Many attempts have been undertaken in order to approximate the
energy eigenvalue spectrum of large magnetic systems. Among
these the Density Matrix Renormalization Group technique (DMRG)
\cite{Whi:PRB93} is one of the most powerful, at least in the
field of one-dimensional spin systems, where for instance
questions concerning Haldane's conjecture \cite{Hal:PL83,Hal:PRL83}
could be answered with great
accuracy.\cite{WhD:PRB93B,GJL:PRB94,Xia:PRB98}
It is as well a powerful tool for studying $T=0$ quantum phase
transitions,\cite{Sch:PRL96} as it delivers accurate results for
ground states. 
The DMRG method was also applied to spin rings -- ``ferric wheels" --
which are quasi one-dimensional magnetic
molecules.\cite{NWZ:PRB01} Here the aim was to evaluate
low-lying 
magnetic levels and the related low-temperature spin dynamics in
order to understand macroscopic quantum coherent phenomena.

Another method of approximating the energy spectrum is
stimulated by the observation that in many Heisenberg 
spin systems the low-lying energy levels $E_{min}(S)$ form a
rotational band,\cite{ScL:PRB01} i.e. they depend approximately
quadratically on the total spin quantum number $S$.
Experimentally this property has been described as ``following
the Land\'{e} interval rule".\cite{TDP:JACS94,LGC:PRB97A,LGC:PRB97B,ACC:ICA00}
For spin ring systems the low-energy spectrum consists of a
sequence of rotational bands, which allows to address questions
of spin tunneling and other transitions without diagonalizing
the full Hamiltonian.\cite{Wal:PRB02}

The purpose of this article is twofold. We show that DMRG can be
used in order to approximate the low-lying energy levels of
magnetic macromolecules like \mofe, and we prove numerically 
that the lowest levels as a function of total spin $S$ form
indeed a rotational band. The latter observation strengthens the
predictions made in Ref.~\onlinecite{LSM:EPL01}.

\section{DMRG technique}

The DMRG technique \cite{Whi:PRB93} became one of the standard
numerical methods for quantum lattice calculations in recent
years.\cite{Pes99} Its basic idea is the reduction of Hilbert space
while focusing on the accuracy of a target state.
For this purpose the system is divided into subunits -- blocks
-- which are represented by reduced sets of basis states. The
size $m$ of the truncated block Hilbert space is a major input
parameter of the method and to a large extent determines its
accuracy. 
The block basis states are derived from a twice as large system
-- superblock -- by first diagonalizing the Hamiltonian on the
superblock, then building a reduced density matrix from the
superblock ground state, and finally diagonalizing the reduced density
matrix.

DMRG is best suited for chain-like structures. Many accurate results
have been achieved by applying DMRG to various (quasi-)one-dimensional
systems.\cite{WhD:PRB93B,GJL:PRB94,Xia:PRB98} The best results
were found for the limit of infinite chains with open boundary
conditions. It is commonly accepted that DMRG reaches maximum
accuracy when it is applied to systems with a small number of
interactions between the blocks, e.g. systems with only nearest-neighbor
interaction and open boundary conditions \cite{Pes99}.

\begin{figure}[ht!]
\begin{center}
\scalebox{0.23}{
\includegraphics[clip]{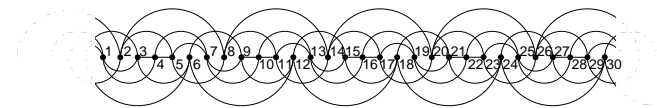}
}
\end{center}
\caption{\label{fe_30_projection_1d.eps}One-dimensional projection of the
icosidodecahedron: the lines represent interactions.}
\end{figure}

In order to apply DMRG calculations to two-dimensional systems a
mapping onto a one-dimensional structure was proposed.\cite{Pes99}
We adopt this idea and derive a simple DMRG algorithm for
two-dimensional spin systems in the Heisenberg model.\cite{Exler:Diplom}
Since the spin array consists of a countable number of spins, any
arbitrary numbering is already a mapping onto a one-dimensional
structure. However, even if the original system had only
nearest-neighbor exchange, the new one-dimensional system has
many  long-range interactions depending on the way the spins are
enumerated, see e.g. Fig.~\ref{fe_30_projection_1d.eps}.
Therefore, a numbering which minimizes long range interactions
is preferable. 

The Hamiltonian of the Heisenberg model, which is appropriate
for the investigated magnetic molecules, can be written as
\begin{equation}
\label{eq:H_J_ij}
\op{H}
=
-
\sum_{i\ne j}{J_{ij}\;\op{\vec{S}}_i \cdot\op{\vec{S}}_j}
=
-2
\sum_{i>j}{J_{ij}\;\op{\vec{S}}_i \cdot\op{\vec{S}}_j}
\eqcomma
\end{equation}
where $J_{ij}$ is the interaction matrix and $\op{\vec{S}}_i$ are the
spin operators at the sites $i$.

\begin{figure}[ht!]
\begin{center}
\scalebox{0.5}{
\includegraphics[clip]{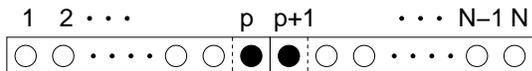}
}
\end{center}
\caption{\label{block_algo_2d.eps}Block setup for second part
  of DMRG algorithm (``sweeping''):
  The whole system of $N$ spins constitutes the
  superblock. The spins $\{1,2,\ldots,p\}$ belong to the left block,
  the other spins $\{p+1,\ldots,N\}$ to the right block.}
\end{figure}
We use a block algorithm similar to the setup in White's original
paper.\cite{Whi:PRB93} Two blocks are connected via two single spin
sites, these four parts form the superblock, see
Fig.~\ref{block_algo_2d.eps}. The Hamiltonian for the superblock
can be written as
\begin{equation}
  \label{eq:H_super}
\op{H}^{\rm SB}
=
\op{H}^{\rm l}+\op{H}^{\rm r}
+
(-2)
\sum_{i=1}^{p}{\sum_{j=p+1}^{N}
{J_{ij}\;\op{\vec{S}}_i\cdot\op{\vec{S}}_j}}\eqcomma
\end{equation}
where $\op{H}^l$ and $\op{H}^r$ represent the Hamiltonians for the left and
right block including the respective single spin. These
Hamiltonians include the interactions inside the respective
blocks, therefore the third term in \fmref{eq:H_super} describes
the interactions of the spins belonging to the left block with
those of the right block.

In our implementation we have to keep track of the operators $\op{S}_i^+$
and $\op{S}_i^z$ (in matrix representation) for all sites $i$. In the case
of a system with pure nearest-neighbor interaction one would have to
keep only the operators for sites at the borders of the blocks.
Because of the long-range interactions, the 2D-DMRG approach
consumes more memory and the calculation of the Hamiltonian
takes more time.

The algorithm consists of two steps.
During the first step the superblock grows with each
iteration by two sites until the final length of the system is reached.
The second step is an implementation of White's
sweep-algorithm.\cite{Whi:PRB93} 
While the superblock is kept at its maximum length, in each
iteration the left block grows by one site whereas the right block
is shortened by one site. 
When the right block reaches the size of two
sites, the direction of the sweep is turned, and the right block
grows in the next iterations. One performs a number of sweeps
until the desired property, in our case the ground state energy,
converges. 

The Hamiltonian is invariant under rotations in spin space.
Therefore the total magnetic quantum number $M$ is a good
quantum number and we can perform our calculation in each
orthogonal subspace \subhm separately.

\section{Accuracy of the method}
\label{sec_acc_DMRG}

Since it is difficult to predict the accuracy of a DMRG calculation,
we apply our implementation to an exactly diagonalizable system
first. The most realistic test system for the use of DMRG for
\mofe\ is the icosidodecahedron with spins $s=1/2$. This fictitious molecule,
which possibly may be synthesized with vanadium ions instead of
iron ions, has the same structure as \mofe, but the smaller spin
quantum number reduces the dimension of the Hilbert space
significantly. Therefore a numerically exact diagonalization is
possible and was carried out by J.~Richter.\cite{SSR:EPJB01,JR:PC}
We use these results to analyze the principle feasibility
and the accuracy of the method.

\begin{figure}
\begin{center}
\scalebox{0.25}{
\includegraphics[clip]{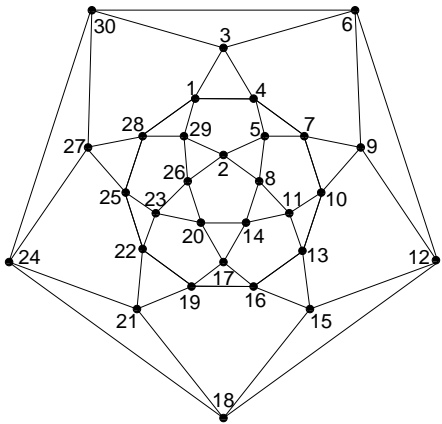}
}
\end{center}
\caption{\label{icosi-graph.eps}Two-dimensional projection of the
icosidodecahedron, the site numbers are those used in our DMRG
algorithm.} 
\end{figure}
The DMRG calculations were implemented using the enumeration of
the spin sites as shown in Figs.~\ref{fe_30_projection_1d.eps}
and~\ref{icosi-graph.eps}.

We choose this enumeration because it minimizes the average
interaction length between two sites.
The interaction length on the chain is measured as the distance
$\left|j-i\right|$ of the two interacting sites $i$ and $j$.
The DMRG method favors systems
with a minimal number of interactions between the blocks.
Therefore, a short average interaction length helps to reduce
the number of inter-block interactions.

With our enumeration we get an average length of $3$ between two
interacting sites. However, the choice is not unique because of the various
symmetries of the system. Our choice,
Fig.~\ref{icosi-graph.eps}, is rotationally symmetric with a
five-fold symmetry corresponding to the five-fold symmetry of
the central pentagon. The sites $1$ to $6$ form the unit cell.

\begin{figure}[ht!]
\begin{center}
\scalebox{0.3}{
\includegraphics[clip]{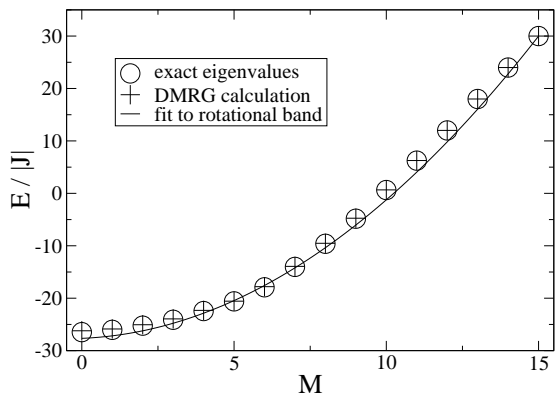}
}
\end{center}
\caption{\label{icosi_S_0_5_rot_band.eps}Eigenvalues and lowest
rotational band of the $s=1/2$ icosidodecahedron.
}
\end{figure}
In Fig.~\ref{icosi_S_0_5_rot_band.eps} our DMRG results
(crosses) are compared to the energy eigenvalues (circles) determined
numerically by J.~Richter with a Lanczos
method.\cite{SSR:EPJB01,JR:PC} We find very good
agreement of both sequences, with a maximal relative error of
about 1\%. Although we don't achieve the high accuracy of
one-dimensional calculations (often better than $10^{-6}$), the
result demonstrates that DMRG is applicable to finite 2D spin 
systems.

Our results were obtained keeping $m=60$ states per block. The number
of states could easily be increased for the $s=1/2$ case, but
we wanted to have a prediction for \mofe, where $m$ is limited
by the available computer resources because of the much larger
spin $s=5/2$. 
The number of sweeps ranged from 5 to 20 depending on how quickly
the algorithm ran into oscillations with no further improvement
of accuracy. This oscillatory behavior of the sweep algorithm is
described in Ref.~\onlinecite{Pes99}.

\section{Rotational band in \mofe}

Since the DMRG technique has proven applicable for the $s=\frac{1}{2}$
case of the icosidodecahedron, we use our algorithm to approximate
energy eigenvalues of the magnetic Keplerate molecule \mofe.

In Ref.~\onlinecite{LSM:EPL01} it was predicted that the low-lying
energy eigenstates of \mofe\ form ``rotational bands'', i.e. the
sequence 
of ground states energies of the sub-spaces \subhs is expected to have a
quadratic dependence on the total spin quantum number $S$. 
A spectrum with rotational bands usually arises in
antiferromagnets if the spin system can be divided into
sub-lattices. The most prominent example are bipartite rings or
chains which consist of two sub-lattices with opposite sub-lattice
magnetization. In the case of \mofe\ the spin system is
decomposable into three sub-lattices with sub-lattice spin quantum
numbers $S_A$, $S_B$, and $S_C$.\cite{ScL:PRB01,LSM:EPL01}
Then the low-lying spectrum can be described by an approximate
Hamilton operator 
\begin{equation}
  \label{eq:H_approx}
  \op{H}_{\rm approx}=-J\frac{D}{N}\left[\op{\vec{S}}^2
    -\gamma\left(\op{\vec{S}}^2_A
      +\op{\vec{S}}^2_B+\op{\vec{S}}^2_C\right)\right]
\eqcomma
\end{equation}
where $\op{\vec{S}}$ is the total spin operator and the others
are sub-lattice spin operators. At least in the case of bipartite
systems this approximation has turned out to be a very good
one.\cite{ScL:PRB01,Wal:PRB02}

The minimal energy eigenvalues of $\op{H}_{\rm approx}$ as a
function of $S$ form a rotational band by construction
\begin{equation}
  \label{eq:E_rot}
  E_{\rm min}\left(S\right)=-J\frac{D}{N}\,S\left(S+1\right)+E_a
\ .
\end{equation}
We use the DMRG method to approximate the lowest energy eigenvalues
of the full Hamiltonian \fmref{eq:H_J_ij} and compare them to
those predicted by the rotational band hypothesis
\fmref{eq:E_rot}. In our calculation we 
obtain energy levels for the ground states of \subhm sub-spaces.
These states are equivalent to the ground states of the sub-spaces
\subhs with $S=M$. The proof for this property rests on the
monotonous increase of the sequence $E\left(M\right)$ with $M$
for $0\le M \le Ns$.

\begin{figure}[ht!]
\begin{center}
\scalebox{0.3}{
\includegraphics[clip]{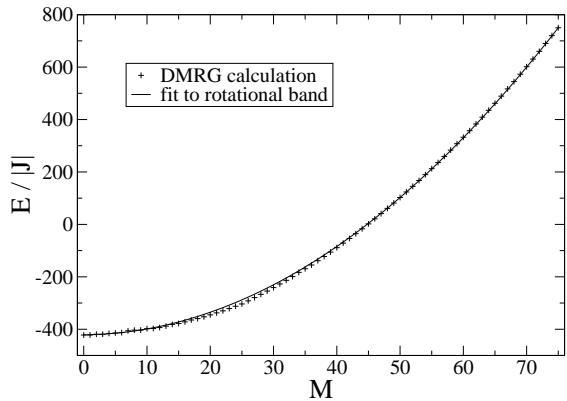}
}
\end{center}
\caption{\label{fe_30_rot_band.eps}DMRG eigenvalues and lowest
rotational band of the $s=5/2$ icosidodecahedron.
}
\end{figure}

Fig.~\ref{fe_30_rot_band.eps} shows our results and a fit to
the lowest rotational band. We find a good agreement between
our DMRG data and the predicted quadratic dependence. From
the fit of our data we obtain $D=6.17$ and $\gamma=1.05$. These
values are very close to the values $D=6.23$ and $\gamma=1.07$
given in Ref.~\onlinecite{LSM:EPL01}, which were inferred from
magnetization measurements.

\section{Summary}

The major result of our investigation is that the DMRG approach
delivers acceptable results for 2D systems as shown in
section~\ref{sec_acc_DMRG} for a fictitious magnetic molecule
of 30 spins $s=1/2$ and icosidodecahedral structure. Therefore,
we assume that our numerical approximation of low-lying energy
levels for \mofe\ is reliable. Thus, we have obtained good
confidence that the prediction of a lowest rotational band
made in Ref.~\onlinecite{LSM:EPL01} is justified. The lowest band
of \mofe\ indeed has a parabolic dependence on $S$. 
It remains
the task of forthcoming investigations whether also the higher
lying rotational bands appear with the same distinctness
or whether they are scattered due to the strong frustration
effects. In any case such calculations demand much higher
precision and thus numerical effort.

The present calculations were carried out keeping $m=60$ block
states, which means that the calculation time for one
ground state is about a day on a standard PC, and there are 
76 \subhm ground states ($M=0,1,\ldots,75$) in \mofe.
For a more accurate calculation one would have to use more
powerful machines and more computer time both allowing higher
values for $m$ and therefore leading to a better accuracy.

\section*{Acknowledgments}

We thank M.~Luban, J.~Richter, and J.~Schulenburg for fruitful
discussions. We also thank the 
National Science Foundation and the Deutscher Akademischer
Austauschdienst for supporting a mutual exchange program with
the Ames Lab.

\bibliography{paper_dmrg}

\end{document}